\documentclass[referee]{raa}
\usepackage{graphicx,times}
\usepackage{natbib}
\usepackage{amssymb,amsmath}
\usepackage{multirow}
\bibpunct{(}{)}{;}{a}{}{,}

\usepackage[a4paper=true,dvipdfm=true,pagebackref=true,unicode]{hyperref}
\hypersetup{pdftitle = The title of my PDF, pdfauthor = My name, pdfsubject= The subject, pdfkeywords = keyword1 keyword2 keyword3}
\hypersetup{colorlinks = true, linkcolor = green, anchorcolor = red, citecolor = blue, filecolor = red, pagecolor = red, urlcolor = red}

\begin{document}

   \title{Sun-Earth connection Event of Super Geomagnetic Storm on March 31, 2001: the Importance of Solar Wind Density}

 \volnopage{ {\bf 20XX} Vol.\ {\bf X} No. {\bf XX}, 000--000}
   \setcounter{page}{1}

   \author{Li-Bin Cheng\inst{1}, Gui-Ming Le\inst{2}, Ming-Xian Zhao\inst{2}
   }

   \institute{School of Physics and Electronic Information,
              Shangrao Normal University,
              Shangrao, 334001, China; \\
          \and
              National Center for Space Weather,
              China Meteorological Administration,
              Beijing, 100081, China ({\it Legm@cma.gov.cn,  zhaomx@cma.gov.cn})\\
\vs \no
   {\small Received 20** June **; accepted 20** July **}
}

\abstract{An X1.7 flare at 10:15 UT and a halo CME with a projected speed of 942 km/s erupted from NOAA solar active region 9393 located at N20W19, were observed on 2001 March 29. When the CME reached the Earth, it triggered a super geomagnetic storm (hereafter super storm). We find that the CME always moved towards the Earth according to the intensity-time profiles of protons with different energies. The solar wind parameters responsible for the main phase of the super storm occurred on March 31, 2001 is analyzed taking into account the delayed geomagnetic effect of solar wind at the L1 point and using the SYM-H index. According to the variation properties of SYM-H index during the main phase of the super storm, the main phase of the super storm is divided into two parts. A comparative study of solar wind parameters responsible for the two parts shows the evidence that the solar wind density plays a significant role in transferring solar wind energy into the magnetosphere, besides the southward magnetic field and solar wind speed.
}

   \authorrunning{L.-B. Cheng G.-M. Le \& M.-X. Zhao}            
   \titlerunning{Sun-Earth connection event on Mar 31 2001}  
   \maketitle

%
\section{Introduction}           
\label{sect:intro}

When a large sunspot or active region (AR) appears near solar disk center, a strong eruption from the region may lead to a severe space weather. If a coronal mass ejection (CME) erupted finally reaches the Earth and then causes a major geomagnetic storm ($Dst \le -100 nT$), this kind of event is called a Sun-Earth connection event. The basic property of a Sun-Earth connection event is that solar atmosphere and interplanetary space are disturbed dramatically by the associated CME, and then a major geomagnetic storm happens. It is generally accepted that if the solar wind has southward magnetic field, the magnetic reconnection between the interplanetary magnetic field and the northward magnetic field of the dayside magnetopause will lead to solar wind energy injection into the Earth's magnetosphere to cause geomagnetic storms. Many papers have been devoted to study the relationship between various solar wind parameters and geomagnetic storm intensities, especially the relationship between peak values of various solar wind parameters and the minimum of $Dst$ indices (e.g., \citealt{Kane+2005, Echer+etal+2008a, Ji+etal+2010, Richardson+2013}). In principle, the occurrence of a geomagnetic storm is due to the sustained magnetic reconnection between solar wind and the magnetosphere. \cite{Tsurutani+Gonzalez+1987} proposed that a major geomagnetic storm would occur, when the southward component of a IMF exceeds 10 nT for 3 hours or longer, or the solar wind electric field exceeds 5 mV/m for more than 3 hours. \cite{Wang+etal+2003b} proposed an empirical formula relating the $Dst$ peak value to solar wind parameters through statistical analysis as follows:
\begin{equation}\label{eq:01}
Dst_{min} = -19.01-8.43\left(\overline{VB_z} \right)^{1.09}(\Delta t)^{0.5}
\end{equation}
where $V$ is the solar wind velocity, $B_z$ is the southward component of IMF, and $\Delta t$ is the time duration. The formula resulted in a correlation coefficient between $Dst$ and $\left(\overline{VB_z} \right)^{1.09}(\Delta t)^{0.5}$ of 0.95. The empirical formula does not take into account the possible contribution made by solar wind density or dynamic pressure. All these studies paid little attention to the possible effect of solar wind density or solar wind dynamic pressure on the intensity of associated storm.

Using global MHD simulations of the solar wind-magnetospheric interaction, \cite{Lopez+etal+2004} have argued that solar wind density may play an important role in the transfer of energy to the magnetosphere during the main phase of a geomagnetic storm. \cite{Kataoka+etal+2005} presented evidence for solar wind density control of energy transfer to the magnetosphere. \cite{Khabarova+and+Yermolaev+2008} believed that solar wind density plays a more significant geoeffective role than velocity. \cite{Weigel+2010} examined the impact of solar wind density on the intensity of a geomagnetic storm, believing that the magnitude of their integrated value can be 1.5 times larger for a given $VB_s$, when the solar wind density is high. However, he cautioned that the role played by the solar wind density would dwindle, when the storm becomes very large, based on the assumption that a large solar wind electric field would go along with a large density.

When a fast CME enters interplanetary space, it becomes an interplanetary coronal mass ejection(ICME). How to trace the ICME propagation in the interplanetary space? If a CME is accompanied by a solar proton event (SPE), the intensity-time profile of SPE can be used to trace the CME propagation in interplanetary space and then to predict the geoeffectiveness of the CME (\citealt{Le+etal+2016, Le+etal+2017, Le+and+Zhang+2017, Zhao+etal+2018}). However, only a small part of intense geomagnetic storms can be predict by using the intensity-time profiles of SPEs. \cite{Smith+etal+2004} \& \cite{Smith+and+Murtagh+2009} proposed that the enhancement of low-energy ions (47-65 keV) observed at L1 point is a potential tool for predicting the arrival of interplanetary shocks hours before they arrive at L1, which can be used to predict large geomagnetic storms. However, which kind of solar wind structures that will follow the enhancement of low-energy ions (47-65 keV) has not been studied, namely that whether the enhancement of low-energy ions (47-65 keV) can be used to trace the propagation of associated ICME in interplanetary (IP) space has not been studied.

Super geomagnetic storm (hereafter super storm) on 2001 March 31 is a typical Sun-Earth connection event. Solar source and interplanetary source have been investigated by \cite{Zhang+etal+2007}. However, the propagation information of the CME from the Sun to the Earth has not been analyzed in the paper of \cite{Zhang+etal+2007}, and whether the relationship between solar wind parameters and the intensities of associated geomagnetic storms described by formula (1) is correct for the super storm of 2001 March 31 has not been studied. Is solar wind density an important parameter for the super storm? How can we know the propagation of the CME from the Sun to the Earth? Whether solar wind density is an important parameter for the super storm? One of the motivation of the present study is to answer these questions.

It is worth noting that the solar wind data in the present study was observed by spacecraft ACE, which is located at L1 point between the Earth and the Sun. As a result, the solar wind at the L1 point can not interact with the magnetosphere immediately only when it reaches there, suggesting that the solar wind at the L1 point would have a delayed geomagnetic effect, though delayed time depending on the solar wind speed.

\cite{Wanliss+and+Showalter+2006} recommended that the SYM-H index can be used as a high-resolution $Dst$ index. As a matter of fact, the SYM-H index has been widely used in geomagnetic storm studies (e.g., \citealt{Nose+etal+2003, Le+etal+2010, Hutchinson+etal+2011}). SYM-H index is applied in this study to describe detailed variation of geomagnetic storm intensity. The main phase of a storm may have multi-dip. Each dip may correspond to a different solar wind structure. The super storm on 2001 March 31 is a singe step storm according to the definition of Dst dip during the main phase of a storm(\citealt{Zhang+etal+2008}). However, the main phase of the super storm is divided into two parts, which are named as step-1 and step-2 respectively, according to the variation speed of SYM-H index during the main phase of the storm in the present study. SYM-H index decreased more quickly during step-1 than during step-2. Various solar wind parameters responsible for step-1 will be compared with those responsible for step-2 to check whether solar wind density is an important parameter for the super storm. In addition, the moving direction of the CME propagating from the Sun to the Earth will also be studied. These are the motivation of the present study. Data analysis are presented in section 2. Discussion and conclusion are presented in final section.

\section{Data analysis}
\label{sect:data}
\subsection{Observations}
\label{sect:obs}

Solar active region (AR) 9393, which is located at N20W19, produced a X1.7 flare at 10:15 UT on 2001 March 29, and a halo CME with projected speed 942 km/s (https://cdaw.gsfc.nasa.gov/CME\_list/) was observed by Large Angle and Spectrometric Coronagraph (LASCO). A solar proton event (SPE) occurred after the eruption of the flare and CME. The shock driven by the CME arrived at L1 point at 00:23 UT and then reached the Earth at 00:52 UT on 2001 March 31. Both the shock and ICME ejecta passed through the Earth and triggered a super storm with the lowest $Dst=-387$ nT on 2001 March 31 (\citealt{Le+etal+2016}).

\subsection{How can we learn the CME propagation in interplanetary space?}
\label{sect:cme}

A solar energetic particle (SEP) event occurred after the eruptions of the CME and flare, which is shown in Figure 1. The flux of $E>10 MeV$ protons increased quickly after the eruptions of the associated flare and CME and then the flux of $E>10 MeV$ protons changed very slowly until the shock reached the Earth. The enhancement in the particle flux heralds the approach of associated interplanetary shock (Smith and Murtagh, 2009). We can see from Figure 1 that fluxes of P1 (47-65 keV) and P8 ($1.88-4.70 MeV$) observed by ACE spacecraft increased with a sustained manner, and the fluxes of the particles in the two channels reached their peak fluxes at the time when shock reached the L1 point. These may suggest that the moving direction of the ICME in interplanetary space is always towards the Earth.

\begin{figure}
   \centering
  \includegraphics[width=10cm, angle=90]{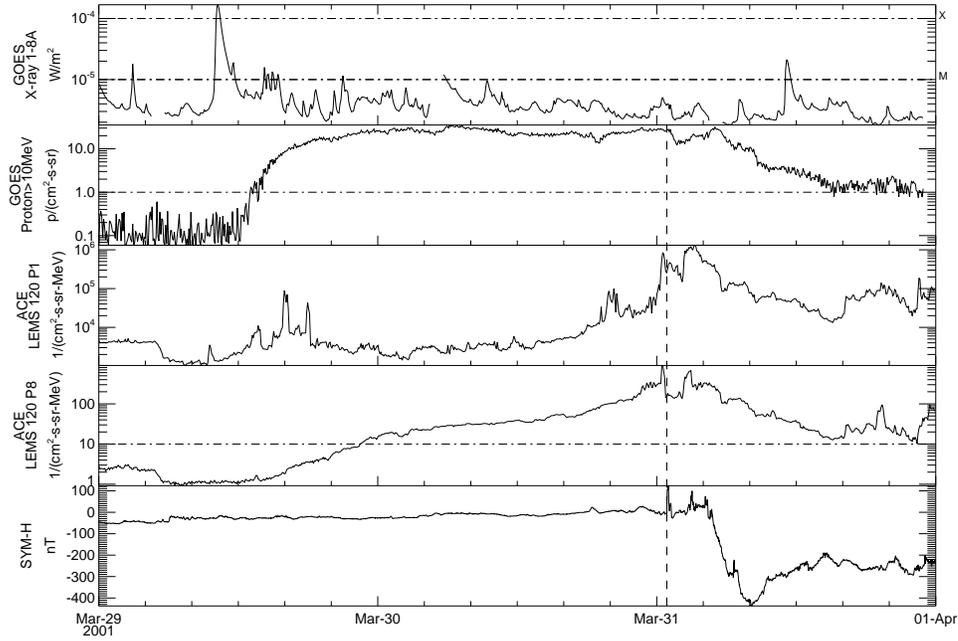}
   \caption{The SEP event associated with the super storm on 2001 March 31. From top to bottom, it shows electromagnetic flux of GOES X-ray 1-8\AA, the particle flux of $E>10 MeV$ protons observed by GOES, the particle flux of P1(47-65 keV) and P8 ($1.88-4.70 MeV$) observed by ACE and SYM-H index. The vertical dashed line indicates the time when the IP shock reached the Earth.}
   \label{fig:01}
\end{figure}

\subsection{The properties of solar wind parameters responsible for the super storm}
\label{sect:solarwind}

To investigate the properties of solar wind parameters responsible for the super storm, the main phase of the super storm, which is constituted by step-1 and step-2, is precisely determined according to the SYM-H index and shown in bottom panel of Figure 2. The preliminary phase is the period between storm sudden commencement(SSC) and the start of the main phase of the storm, which is also shown in bottom panel of Figure 2. Solar wind at L1 point observed by ACE spacecraft can not have an effect on the geomagnetic field immediately, only when it propagates to the Earth. According to the solar wind speed observed by ACE spacecraft on 2001 March 31, solar wind responsible for preliminary phase, step-1 and step-2 of the super storm are period 1, period 2 and period 3 respectively. The start and end time for period 1 and preliminary step, period 2 and step-1, and period 3 and step-2 have been listed in Table 1.

\begin{figure}
   \centering
  \includegraphics[width=12cm, angle=0]{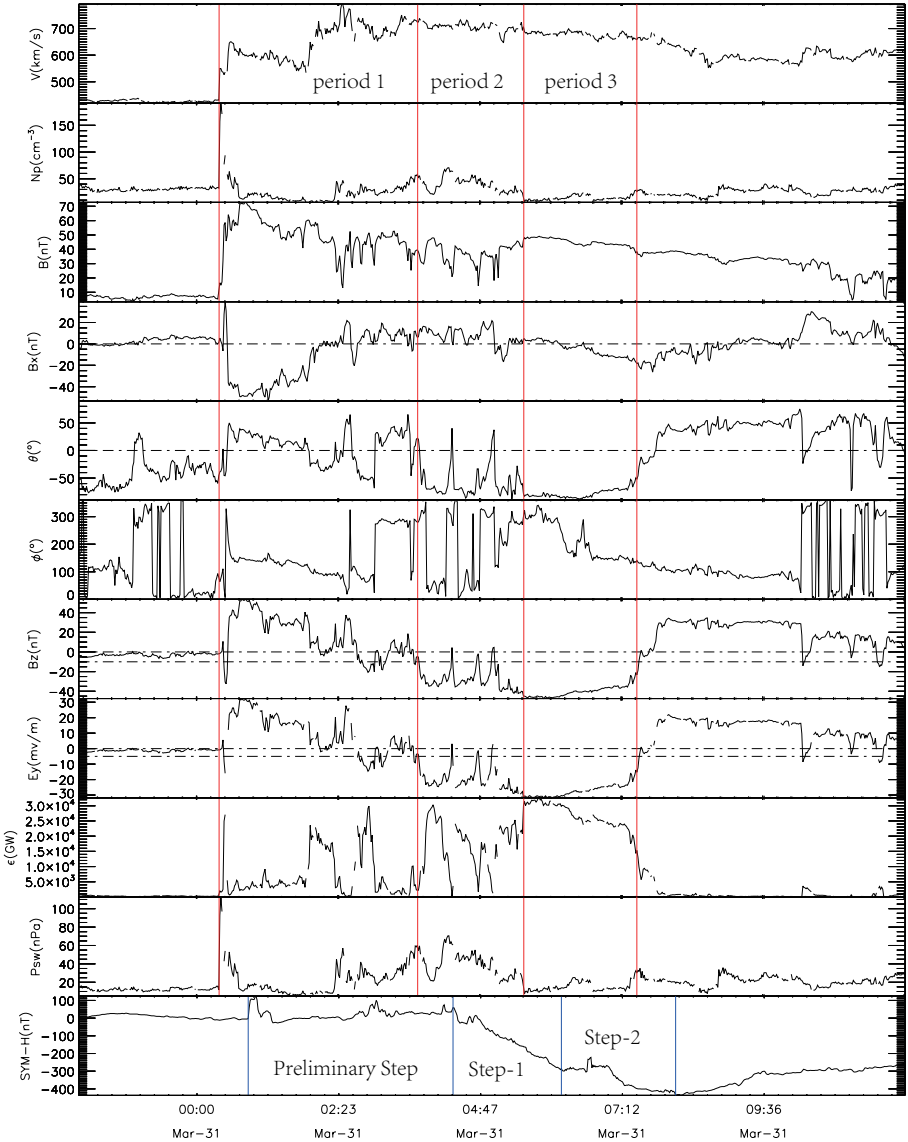}
   \caption{solar wind parameters observed by the satellite ACE and geomagnetic index SYM-H from 20:00 UT March 30 to 13:00 UT March 31, 2001. From top to bottom, it shows solar wind speed ($V$), proton density ($Np$), IMF strength ($B$), the elevation $\theta$ and azimuthal $\phi$ angles of IMF direction, z-component field of IMF($B_z$), solar wind electric field ($E_y$), Akasofu energy coupling function ($\varepsilon$),  solar wind dynamic pressure ($P_{sw}$,) and geomagnetic index SYM-H. The first red vertical solid line indicates the arrival of IP shock. The second red vertical line indicates the time 03:46UT. The third and fourth red vertical solid lines indicate the moment 05:25 UT and 07:28 UT respectively. The first blue vertical solid line indicates the SSC. The blue vertical dot dashed line indicates the time 04:20UT. The second and third blue vertical solid lines indicate the moment 06:00 UT and 08:06 UT respectively}
   \label{fig:02}
\end{figure}

\begin{table}
\bc
\begin{minipage}[]{110mm}
\caption[]{The time comparison between solar wind and the SYM-H index\label{tab:01}}\end{minipage}
\setlength{\tabcolsep}{1pt}
\small
 \begin{tabular}{|c|c|c|c|}
  \hline
  ~ & Period 1 & Period 2 & Period 3 \\
  ~Time of Solar wind at L1 point~ & ~(00:23-03:46 UT)~ & ~(03:46-05:25 UT)~ & ~(05:25-07:28 UT)~ \\
  \hline
  ~ & Preliminary step & Step-1 & Step-2 \\
  Time of SYM-H & ~(00:59-04:20 UT)~ & ~(04:20-06:00 UT)~ & ~(06:00-08:06 UT)~ \\
  \hline
\end{tabular}
\ec
\end{table}

We focus on the solar wind parameters that are responsible for the main phase of the super storm. The variation of SYM-H index during step-1 is $-319$ nT, while the variation of SYM-H index during step-2 is $-177 nT$, much smaller than that during step-1. Time duration of step-1 is 100 minutes, while time duration of step-2 is 126 minutes, which is longer than that of step-1. It is evident that the variation of SYM-H index during step-1 is much more dramatically than that during step-2, and step-1 made much more contribution to super storm than step-2.

\begin{table}
\bc
\begin{minipage}[]{110mm}
\caption[]{The variations of SYM-H during Step-1 and Step-2\label{tab:02}}\end{minipage}
\setlength{\tabcolsep}{1pt}
\small
 \begin{tabular}{|c|c|c|c|c|}
  \hline
  ~Periods~ & \multicolumn{2}{|c|}{Step-1} & \multicolumn{2}{|c|}{Step-2} \\
  \hline
  \multirow{2}{*}{ } & ~Start time~ & ~End time~ & ~Start time~ & ~End time~ \\
  \cline{2-5}
  \multirow{2}{*}{ } & ~04:20 UT~ & ~06:00 UT~ & ~06:00 UT~ & ~08:06 UT~ \\
  \hline
  ~$\Delta t$~(min)~ & \multicolumn{2}{|c|}{$100$} & \multicolumn{2}{|c|}{$126$} \\
  \hline
  ~$\Delta \text{SYM-H}$ (nT)~ & \multicolumn{2}{|c|}{$-319$} & \multicolumn{2}{|c|}{$-177$} \\
  \hline
\end{tabular}
\ec
\end{table}

Solar wind energy coupling function, $\varepsilon$, proposed by \cite{Akasofu+1981} is calculated by the formula listed below:
\begin{equation}\label{eq:02}
\varepsilon = VB^2 sin^4 \left(\theta/2\right)l_0^4
\end{equation}
where $V$,$B$,$\theta$ represent solar wind speed, magnetic field, and the polar angle of the magnetic field vector projected onto the Y-Z plane, respectively. $l_0$ means 7 times the Earth's radius.

The Burton equation has the form (\citealt{Burton+etal+1975}),
\begin{eqnarray}
\mathrm{d}Dst^*/\mathrm{d}t=Q(t)-Dst^*/\tau \label{eq:03}\\
or~~~ \mathrm{d}\text{SYM-H}^*/\mathrm{d} t=Q(t)-\text{SYM-H}^*/\tau \label{eq:05}
\end{eqnarray}
Where $Dst^*$ or $\text{SYM-H}^*$ is the pressure-corrected $Dst$ or SYM-H index and the contribution made by the magnetopause current has been subtracted in (2) and (3) (\citealt{OBrien+and+McPherron+2000}). $\tau$ and $Q$ are the decay time and the injection term of the ring current, respectively. $Q$ has the following form (\citealt{Wang+etal+2003a}):
\begin{equation}\label{eq:05}
  Q = \left\{ \begin{array}{ll}
                -4.4(VB_s-0.49)(P_{sw}/3)^{0.5} & VB_s > 0.49 mV/m \\
                0 & VB_s \le 0.49 mV/m
              \end{array} \right.
\end{equation}
where, $V$ is the solar wind speed, $B_s$ is the southward component of IMF and $P_{sw}$ is solar wind dynamic pressure.

To calculate various solar wind parameters, solar wind data with 64s time resolution observed by ACE spacecraft is used in the study. Various solar wind parameters responsible for step-1 and step-2 have been calculated and listed in Table 3.

\begin{table}
\bc
\begin{minipage}[]{110mm}
\caption[]{Various solar wind parameters during period 2 and period 3\label{tab:03}}\end{minipage}
\setlength{\tabcolsep}{1pt}
\small
 \begin{tabular}{ccc}
  \hline
  ~Parameters~ & ~Period 2~          & ~Period 3~\\
  ~            & ~(03:46-05:25UT)~   & ~(05:25-07:28 UT)~ \\
  ~$\Delta t$~ & ~1h39min~           & ~2h3min~ \\
  ~$\int_{ts}^{te}B_s\mathrm{d}t ~(nT \cdot min)$~ & ~$-2572.65$~ & ~$-4640.5$~ \\
  ~$\overline{B_z}~(nT)$~ & ~$-27.9$~ & ~$-40$~ \\
  ~$\int_{ts}^{te}E_y\mathrm{d}t ~ (mV/m \cdot min)$~ & ~$-1811.2$~ & ~$-3137.5$~ \\
  ~$\overline{E_y}~(mV/m)$~ & ~$-19.7$~ & ~$-27$~ \\
  ~$\int_{ts}^{te}\varepsilon \mathrm{d}t~(GW)$~ & ~$1.6\times10^6$~ & ~$3.05\times10^6$~ \\
  ~$\overline{\varepsilon}~(GW/m)$~ & ~$17379.6$~ & ~$26339.2$~ \\
  ~$\overline{N_p}~(1/cm^3)$~ & ~$40.4$~ & ~$14.9$~ \\
  ~$\overline{P_d}~(nPa)$~ & ~$33.6$~ & ~$11.6$~ \\
  ~$\overline{\theta}~(\deg)$~ & ~$59.2$~ & ~$76.1$~ \\
  ~$\overline{Q}~(nT/min)$~ & ~$-280.9$~ & ~$-222.1$~ \\
  \hline
\end{tabular}
\ec
\end{table}

We can see from Table 3 that solar wind during period 3 has larger $B_z$ and $E_y$ than solar wind during period 2. Time integral of $B_z$ and $E_y$ during period 3 are also larger than those during period 2. However, the contribution to the main phase of the super storm made by solar wind during period 2 is much larger than that made by solar wind during period 3. We can see from Table 3 that averaged solar wind dynamic pressure during period 2 is much larger than that during period 3. This leads to that the dynamic pressure and averaged injection function, $\overline{Q}$ , during period 2 is much larger than those during period 3. This may be the reason that solar wind during period 2 made much more contribution to the super storm than solar wind during period 3, implying that solar wind density is an important parameter for the development of super storm.

\section{Discussion and Conclussion}
\label{sect:discussion}

Because of low energy of P1, weak shock, which is driven by far flank of the associated CME, may also lead to the enhancement in the flux of P1. In this context, only the enhancement in the flux of P1 can not ensure that both the associated IP shock and ejecta will pass the Earth. Here, we give an example shown in Figure 3. The CME associated with the solar proton event that began on 2001 January 28 shown in Figure 2 in the paper of \cite{Le+etal+2016} missed the Earth, only the shock driven by far flank of associated ICME crossed the Earth. We can see from Figure 3 that the flux of P1 increased with a sustained manner and reached its peak flux at the moment when the shock reached L1 point. However, the flux of P8 increased quickly at the early phase and reached its peak flux no long after, and then the flux of P8 declined gradually. The flux of P8 still declined at the time when IP shock passed the ACE spacecraft, indicating that the IP shock can not accelerated particles with energies (1.88 - 4.70 MeV ) efficiently, namely that the shock is really a weak shock. Because only the far flank shock passed the Earth, only a small magnetic storm followed after the peak flux of P1. The comparison between Figure 1 and Figure 3 suggests that only the sustained enhancement in the flux of P1 can not ensure that both associated IP shock and ICME will pass the Earth. However, the phenomena that the fluxes of both P1 and P8 increase with a sustained manner and reach their peak fluxes at the time when associated IP shock passes ACE spacecraft may imply that both IP shock and the associated ICME will pass the Earth, namely that the associated ICME may always moves towards the Earth. Statistical study will be made in the near future.

\begin{figure}
   \centering
  \includegraphics[width=10cm, angle=90]{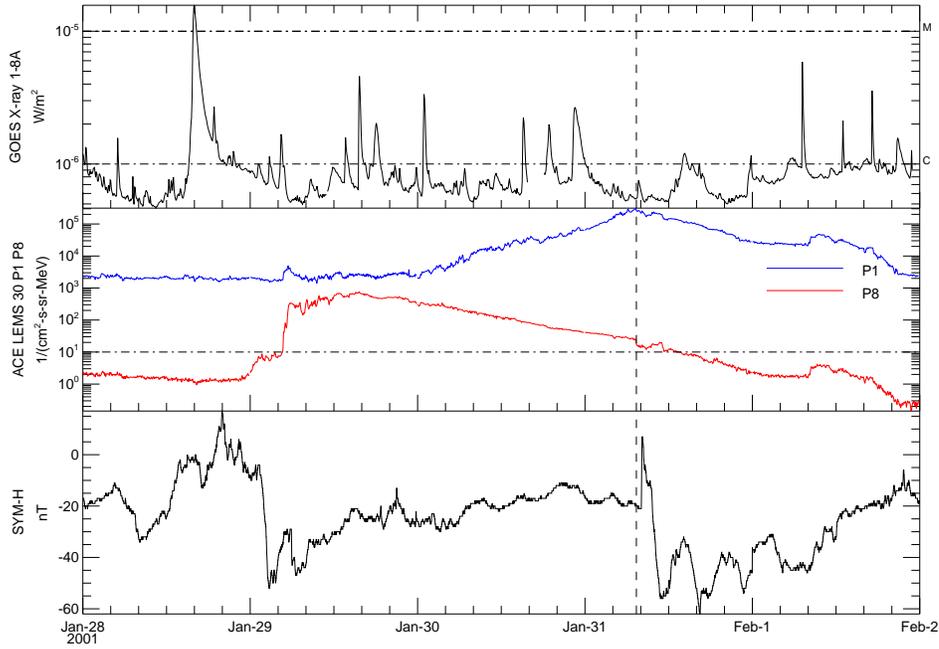}
   \caption{The SEP event associated with the super storm on 2001 January 31. From top to bottom, it shows flux of $1-8\AA$, the flux of P1(blue line) and P8 (red line) observed by ACE and SYM-H index. Vertical dashed line indicates the arrival of the IP shock observed by ACE at 07:22 UT, 2001 January 31.}
   \label{fig:03}
\end{figure}

According to the empirical formula (1) established by \cite{Wang+etal+2003b}, the solar wind during Period 2 would resulted in $Dst_{min}=-279$ nT. However, the real variation of $Dst$ caused by solar wind during period 2 is $-319$ nT. In the same manner, the variation of $Dst$ caused by solar wind during Period 3 should be $- 425$ nT according to formula (1). However, the real variation of $Dst$ caused by solar wind during period 3 is $-177$ nT. Apparently, the variation of $Dst$ estimated by formula (1) is not correct for the present super storm.

Period 3 has longer duration than period 2, and many solar wind parameters during period 3 are larger than those during period 2 except solar wind density. However, solar wind during period 2 made much more contribution to the main phase of the super storm than solar wind during period 3. These suggest that solar wind density or solar wind dynamic is also an important parameter for the super storm, which supports the conclusion obtained by the two papers of \cite{Kataoka+etal+2005} and \cite{Weigel+2010}.

Two CMEs erupted on 2001 March 28 and one CME erupted on 2001 March 29. They finally reached the Earth and formed multiple clouds (\citealt{Wang+etal+2003c}). Two magnetic clouds or two-ejecta associated the super storm observed by ACE on 2001 March 31 have been studied by some researchers (e.g., \citealt{Wang+etal+2003c, Farrugia+etal+2006}). All those papers focused on the influence of CME-interaction on the super storm, which is described by $Dst$ index. The present study focus on the quick and detailed variation of ring current during the main phase of the super storm, which can only be described by SYM-H index, and the influence of solar wind density on the super storm. It is worthy noting that solar wind density is not an independent parameter. Solar wind density always works together with solar wind speed and magnetic field. If solar wind density is not considered, the correlation coefficients between $Dst$ and both south component magnetic field ($B_s$) and solar wind electric field ($E_y$) are very low, and the correlation coefficient between $Dst$ and the time integrated solar wind $E_y$ parameter is only 0.62 (\citealt{Echer+etal+2008b}). A statistical study of the influence of solar wind density on major geomagnetic storm intensity will be made in the near future.

The present study has led to the following conclusions:

The intensity-time profiles of the particles with different energies associated with the halo CME with a projected speed 942 km/s erupted from AR 9393 on 2001 March 29 imply that the CME may always move towards the Earth. The comparison of solar wind parameters responsible for the two different parts of the main phase of the super storm shows the evidence that solar wind density plays a significant role in transferring solar wind energy into the magnetosphere, besides the southward magnetic field and solar wind speed.
\normalem
\begin{acknowledgements}

We are very grateful to the anonymous referee for her/his reviewing of the paper carefully and for helpful suggestions. We thank NOAA for providing the solar soft X-ray and SPE data, and the Data Analysis Center for Geomagnetism and Space Magnetism, Kyoto University, for providing the $Dst$ and SYM-H indexes. We would like to thank the ACE SWEPAM instrument team and the ACE Science Center for providing ACE data. We also thank Institute of Geophysics, China Earthquake Administration for providing sudden storm commence data. This work is supported by the National Natural Science Foundation of China (Grant No. 41074132, 41274193, 41474166).

\end{acknowledgements}

\end{document}